# Analysis of Frequency-Diverse and Dispersion Effects in Dynamic Metasurface Antenna for Holographic Sensing and Imaging

*Abdul Jabbar, Aakash Bansal, and William Whittow

Wolfson School of Mechanical, Electrical and Manufacturing Engineering, Loughborough University, UK

*a.jabbar@lboro.ac.uk

*Abstract*—Dynamic metasurface antennas (DMAs) represent a novel approach to programmable and affordable electromagnetic wave manipulation for enhanced wireless communications, sensing, and imaging applications. Nevertheless, current DMA designs and models are usually quasi-narrowband, neglecting the versatile frequency-diverse manifestation and its utilization. This work demonstrates the frequency-diversity and dispersion operations of a representative DMA structure at the millimeter-wave band. We demonstrate flexible dispersion manipulation through dynamic holographic reconfigurability of the meta-atoms in a DMA. This effect can create distinct radiation patterns across the operating frequency band, achieving flexible frequency diversity with enhanced scanning range within a compact, reconfigurable platform. It eliminates the need for wideband systems or complex phase-shifting networks while offering an alternative to frequency-scanned static beams of traditional leaky-wave antennas. The results establish fundamental insights into modelling and utilization of dispersive effects of DMAs in next-generation near-field and far-field holographic sensing and computational holographic imaging applications.

## I. Introduction

Radio-frequency (RF) sensing and imaging have gained huge interest over the past two decades due to the rapid advancement of mobile and wireless technologies [1], [2]. The evolution from traditional communication-centric systems toward multifunctional platforms has enabled RF hardware and signal processing paradigms to serve not only as information carriers but also as powerful tools for environmental perception and awareness.

Dynamic metasurface antenna (DMA) is an innovative antenna technology that enables versatile control of electromagnetic (EM) waves through the tunable resonance behavior of its constituent unit cells (meta-atoms), offering new possibilities for wireless applications. In the past, phased array antennas, frequency diverse antenna arrays, and metasurface apertures have been utilized for RF sensing and imaging applications. Before discussing the DMA-based frequency-diversity and dispersion operation, a brief overview of related antenna technologies is presented below.

### A. Spatially Diverse Phased-array Antennas

Phased arrays and electronically scanned antennas leverage electronic beamsteering and sample a scene using a densely populated aperture of antennas. Phased array and synthetic aperture radars have been widely utilized for decades for RF radar sensing and imaging applications [3], [4]. Nevertheless, these legacy systems offering spatial diversity suffer from exorbitant challenges such as reduced speed of operation and high cost in both hardware and signal processing [5]. Considering that phased arrays rely on high-resolution phase shifter integrated circuits (ICs) and complex corporate feed networks, the hardware cost and power consumption will become unaffordable, especially in millimeter-wave (mmWave) systems [6].

### B. Frequency Diverse Antenna Arrays

Frequency diversity is another pervasive technique for RF sensing and imaging [7]. Leaky-wave antennas (LWAs) have been extensively reported to produce frequency-scanning beams due to the series-feed network. These antennas can have a limited scanning range in a narrow frequency band, due to which they are required to offer a wideband operation to extend the frequency-dependent scanning range [8], [9]. Likewise, various high-scanning-rate LWAs configurations have been proposed to reduce bandwidth requirements for frequency-diverse operations [10]–[13]. These antennas can achieve RF data collection by sweeping frequency without mechanical moving parts or active switching of phase shifters. However, LWAs inherently lack the reconfigurability of the beam once they are designed, offering a limited degree of control over their radiation behavior. It is instructive to mention here that the frequency beamscanning of LWAs is different than frequency diverse arrays. In contrast to conventional phased arrays, whose beam pointing direction remains fixed once the spatial phase distribution is defined, frequency diverse arrays introduce a small frequency offset between adjacent radiating meta-atoms, producing a time-varying interference pattern that couples range, angle, and time [14]–[18]. This joint aperture-waveform design enables dynamic beam focusing rather than a static far-field beam. However. Such frequency diverse arrays require multiple RF chains and active frequency offset control that increases hardware complexity.

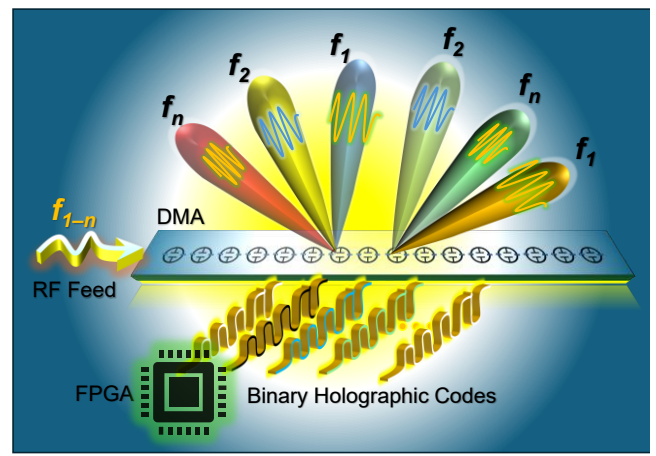

Fig. 1. Conceptual illustration of frequency-diversity through DMA.

### C. Frequency Diverse Metasurface Apertures

Metasurface-based radiative apertures have been investigated extensively to produce frequency diversity to generate pseudorandomly distributed electric fields at different frequencies [19]–[23]. These surfaces consist of randomly distributed slots or irises, and the aperture is excited through a single feed that is launched inside the surface and traverses through the structure, and radiates through the irises [24], [25]. These systems encode the scene using a series of low-correlated radiation patterns generated by the metasurface, enabling high-resolution sensing and image reconstruction through computational post-processing [26], [27]. The performance of metasurface-based structures depends critically on the orthogonality and diversity of the radiation modes produced by the antenna. However, such frequency-diverse sensing architectures inherently demand a relatively wide operating bandwidth, which leads to inefficient spectral utilization, high analog-to-digital sampling rates, and degraded signal-to-noise ratio, specifically at mmWave bands. Moreover, the beam patterns are often random with high side lobe levels in such apertures, which can reduce sensing and imaging resolution.

### D. Dynamic Metasurface Antennas

DMA (also known as Reconfigurable Holographic Surface) is an innovative antenna technology that provides diverse EM wave control features and pattern diversity through a simplified hardware platform [28], [29] It offers radiation pattern diversity through digital control of its constituent meta-atoms without using phase shifters and a complex feed network. In recent years, DMAs have been reported to be utilized for communication [30]–[32], sensing [33], [34], and imaging [35]–[38] applications using their software-controlled pattern diversity feature. However, these applications mainly leverage DMA's pattern diversity. The diverse dispersive paradigm and frequency diversity of DMAs have remained an untapped area of research that needs explicit exploration to avail further benefits, as illustrated in Fig. 1.

It is instructive to mention here that the design and construction of the DMAs/RHSs share some common features with those of the LWAs; however, traditional LWAs only radiate through slots, and hence their EM response cannot be reconfigured once they are designed. In contrast, the essence of DMAs is to use subwavelength-sized meta-atoms, which can dynamically tune the guided wave mode into any desired free-space EM mode. Hence, DMAs can form not only spatially varying EM responses but also offer advanced temporal control through programmable digital control.

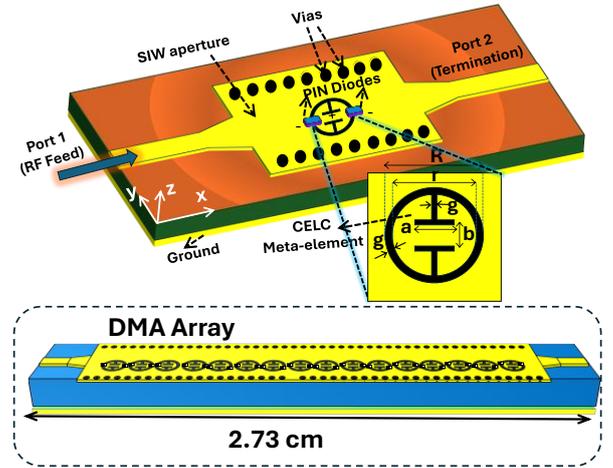

Fig. 2. Schematic view of the DMA meta-atom and array. Dimensions in mm ($a$ = 0.3, $b$ = 0.2, $g$ = 0.1, $R$ = 1.1, $r$ = 0.9).

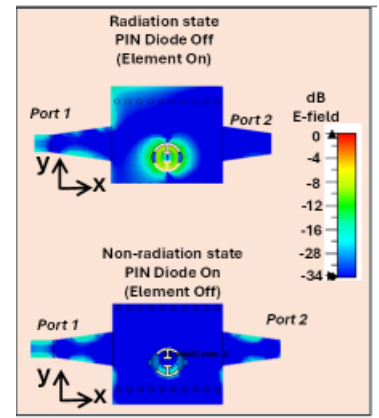

Fig. 3. Radiation profile of CELC meta-atom during on and off states.

In this paper, we present full-wave EM results of a mmWave substrate-integrated waveguide (SIW)-fed dynamic metasurface antenna (DMA) that demonstrates dispersive characteristics and frequency-diverse radiation patterns. The dispersive characteristics of the constituent meta-atoms are analyzed to reveal their influence on the overall aperture response. The results demonstrate that the proposed DMA can operate as a frequency-diverse beam-scanning array, offering a programmable substitute for traditional LWAs. Furthermore, it is shown that the radiated patterns vary not only with the limited available excitation frequency band but also with the applied binary holographic codes, revealing the potential of DMAs as versatile multi-functional metasurface-controlled antenna systems for future adaptive RF sensing and imaging platforms.

## II. Dispersion Analysis and Design of DMA

### A. Design Principle and Holographic Coding Approach of DMA

The DMA is based on a V-band edge-fed SIW structure (substrate is RO3003 with dielectric constant of 3 and loss tangent of 0.001) that excites a periodic array of sub-wavelength complementary electric-inductive-capacitive (CELC) meta-atoms, etched on the top metallic layer, as shown in Fig. 2. The CELC meta-atom is designed to provide a tunable Lorentzian resonance response through electronic bias control of attached PIN diodes, controlling its radiation state (0 or 1). Hence, the CELC can be dynamically activated or deactivated. The detailed geometrical dimensions are provided in our previous work [28]. However, the frequency-diverse phenomenon of DMA presented here is novel and has not been reported previously. As shown in Fig. 3, the CELC meta-atom can be switched on and off to control the radiation state through biasing control of PIN diodes. The scattering parameters of the DMA CELC meta-atom in radiation and non-radiation states are shown in Fig. 4(a).

For the linear 16-element DMA array, since the activated meta-atoms radiate part of the waveguide's energy to free space, the overall radiation pattern of the DMA stems from the superposition of all the activated meta-atoms. Their excitation states are tunable by a binary coding sequence of integrated PIN diodes controlled through a field programmable gate array (FPGA). Based on the holographic interference principle, each DMA element can control the radiation amplitude of the incident EM waves electronically to generate object beams, and such a beamforming technology is also known as holographic beamforming. Therefore, by tuning the EM characteristics of the meta-atoms and enabling/disabling different radiating sets, the DMA supports dynamic radiation pattern diversity in a software-controllable manner at both fixed and varying frequencies, realizing a compact, low-cost, and reconfigurable platform.

### B. Dispersive Effects and Flexible Dispersion Control

Around the resonant frequency, the CELC meta-atom exhibits anomalous dispersion, where the effective permittivity, refractive index, and group velocity attain negative values. Anomalous dispersion is characterized by a negative derivative of the refractive index with respect to frequency ($dn/d\omega<0$), as evident in Fig. 4(d). Such behaviour is absent in conventional materials. Since the CELC meta-atom behaves as a magnetic dipole, it exhibits Lorentzian-type dispersive behaviour and couples effectively with the underlying waveguide mode. This interaction produces a resonance profile characterized by a dip in the transmission magnitude around 60 GHz, indicating strong dispersive characteristics in the resonant state. In addition, the steep phase variation of the transmission coefficient near resonance results in a double phase-slope reversal, as depicted in Fig. 4(b). The positive slope of the transmission phase ($\phi_{S_{21}}$) with respect to angular frequency ($\omega$) leads to a negative group delay $\tau_g$, related as: $\tau_g = -\frac{d\phi_{S_{21}}}{d\omega}$. Moreover, the group velocity $v_g$ and group refractive index $n$ are related as $v_g = \frac{c}{n(\omega) + \omega(\frac{dn}{d\omega})}$. This in turn corresponds to a negative group velocity and negative group refractive index, as shown in Fig. 4(c) and (d), respectively. These effects collectively indicate that the energy front of the radiated wave experiences an opposite temporal shift within the resonant band, confirming the presence of a strongly dispersive, slow-wave regime in the DMA aperture that can be harnessed for frequency-diverse radiation control and wave–matter interaction enhancement.

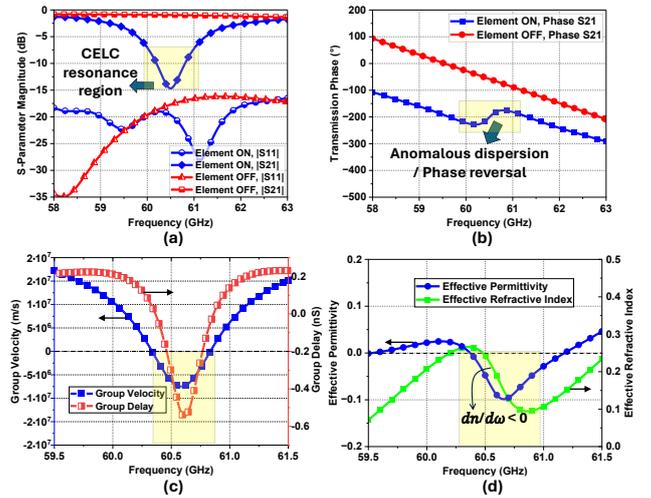

Fig. 4. (a) S-parameters of the DMA meta-atom. (b) transmission phase. (c) dispersive indicators: group delay and group velocity. (d) group refractive index, and effective permittivity of the meta-atom.

The dispersive effects and frequency diversity of DMA can be made flexible through applied binary coding sequences. Flexible dispersion manipulation is critical for holographic applications to achieve advanced imaging or frequency division multiplexing. For each applied coding sequence, the local surface impedance and resonance state of the meta-atoms can be dynamically varied. In this way, we can illustrate the unprecedented control over dispersion characteristics of the DMA across the frequency spectrum without the need for additional phase-shifting hardware.

## III. Frequency Diverse Beam Control and Results

To validate the frequency-diverse behavior of the DMA, the radiation patterns were simulated at 60 GHz, 61 GHz, and 62 GHz for various binary holographic sequences applied across the aperture. The scattering parameters and radiated power for two such coding sequences are presented in Fig. 5, covering effectively the 59–63 GHz band for imaging and sensing applications, where the $S_{11}$ matching level can be less stringent than that of the communication domain.

The beam patterns across frequency for six distinct coding configurations are shown in Fig. 6. Each digital coding sequence can be interpreted as a binary hologram, defining a spatial impedance distribution across the metasurface aperture that governs the radiated wavefront and beam direction. We observed that the beam direction changes with the frequency for each fixed binary hologram. This verifies that the DMA exhibits frequency-diverse beam variation, even under fixed bias conditions, arising from the collective phase response of the meta-atoms.

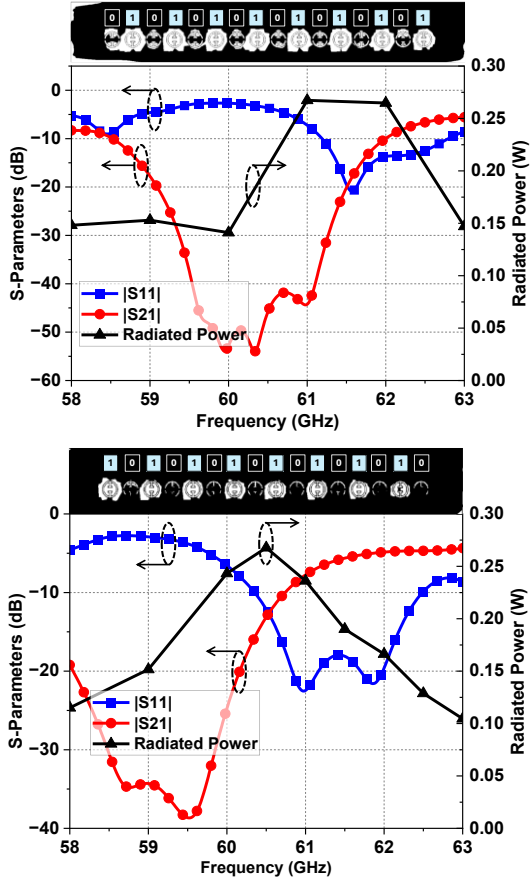

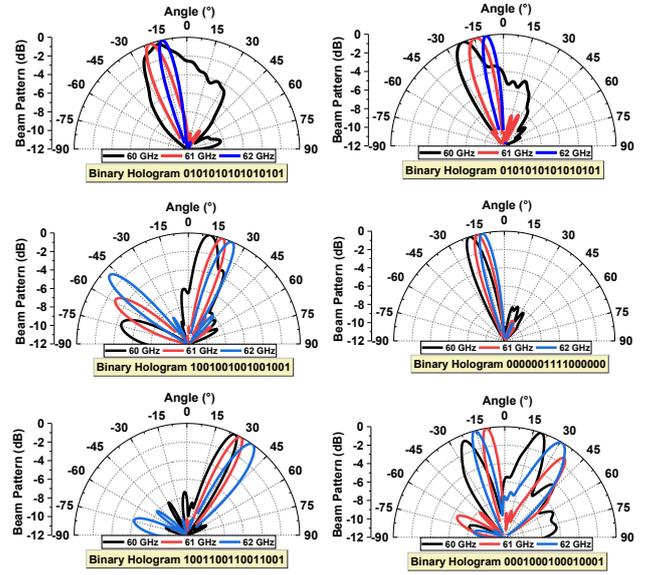

Fig. 6. Frequency-diverse beam patterns of DMA for dynamic holograms.

Fig. 5. Scattering parameters and radiation power profile of the DMA for different binary holograms.

The scan range can be further enhanced and dynamically changed through the application of different binary holograms. In addition, note that it not only produces the frequency diversity but also changes the beamwidth of the beams across the frequency. Hence, DMA can illuminate a scene with dynamic patterns (where each configuration can be called a "mask") that can be altered based on the dynamic holographic sequences. Such dynamic flexibility in beam shaping is a promising advantage for a multitude of imaging and sensing applications, according to the size profile of the subject under test.

Furthermore, it is worth mentioning here that, unlike LWAs and frequency diverse arrays, where the beam direction is governed solely by the operating frequency, or complex phase shifting in the path of feed lines, the DMA introduces an additional degree of freedom through programmable holographic coding. Each binary holographic configuration defines a distinct surface impedance distribution, resulting in advanced dispersion control and producing a unique radiation pattern that further evolves across the operating frequency band. Consequently, while the frequency-induced scan range may be limited, the combination of frequency dispersion and digital reconfigurability enables a versatile set of spatial modes, effectively realizing hybrid frequency-code diversity. This hybrid frequency-code diversity controllable diversity allows the DMA to emulate the wideband spatial adaptability of traditional frequency diverse arrays, but with significantly lower hardware complexity, energy consumption, as well as without requiring wideband feed networks and phase-shifting complexities in the RF chain or RF feed paths. These results confirm that a DMA can serve as a compact and energy-efficient platform supporting reconfigurable frequency-diverse beam scanning and pattern diversity, for computational imaging, and adaptive mmWave sensing applications.

IV. CONCLUSION

This work presented the dispersive effects and frequency-diverse functioning of a DMA at the 60 GHz mmWave band. The dispersive properties of the constituent meta-atoms of DMA are elucidated. Advanced dispersion control of DMA is demonstrated through applied holographic binary coding. The findings show that the tunable Lorentzian resonance response of the meta-atoms can be exploited to achieve flexible frequency-diverse beam control, lowering the dependence on hardware complexity, and wideband antenna networks. It also eliminates the requirements of expensive high-frequency phase-shifting complexities in the RF chain or RF feed paths. Additionally, the versatile programmability of the DMA offers flexible dispersion manipulation, as well as a low-cost means of achieving controllable spatial–spectral responses. This holds the potential to enable high-resolution mmWave RF sensing, and dynamic computational imaging.